\documentclass{ws-procs975x65}

\begin{document}

\title{A TRANSPORT EQUATION APPROACH TO GREEN FUNCTIONS AND SELF-FORCE CALCULATIONS}

\author{Barry Wardell$^{1,2,*}$ and Adrian C. Ottewill$^{1,\dagger}$}
\address{$^1$Complex and Adaptive Systems Laboratory and School of Mathematical Sciences, \\
University College Dublin,
Belfield, Dublin 4, Ireland\\
$^2$Max-Planck-Institut f\"ur Gravitationsphysik, Albert-Einstein-Institut,
14476 Potsdam, Germany\\
$^*$E-mail: barry.wardell@aei.mpg.de\\
$^\dagger$E-mail: adrian.ottewill@ucd.ie
}

\begin{abstract}
In a recent work, we presented the first application of the Poisson-Wiseman-Anderson method of `matched expansions' to compute the self-force acting on a point particle moving in a curved spacetime. The method employs two expansions for the Green function which are respectively valid in the `quasilocal' and `distant past' regimes, and which may be matched together within the normal neighbourhood. In this article, we introduce the method of matched expansions and discuss transport equation methods for the calculation of the Green function in the quasilocal region. These methods allow the Green function to be evaluated throughout the normal neighborhood and are also relevant to a broad range of problems from radiation reaction to quantum field theory in curved spacetime and quantum gravity.
\end{abstract}

\keywords{Self-Force; Green Function; Transport Equation}

\bodymatter

\section{Matched Expansion Approach to the Self-Force}
A charged particle moving in a curved background spacetime will not exactly follow a geodesic of that spacetime. Within the framework of perturbation theory, the particle is seen at linear order to exert a force on itself, a \emph{self-force}, which is responsible for the particle moving away from geodesic motion.

A formal expression for this self-force has been derived \cite{DeWitt:1960,Mino:Sasaki:Tanaka:1996,Quinn:Wald:1997,Quinn:2000} in terms of some easily calculated local terms and a \emph{tail} integral over the entire past-history of the particle. For the case of a scalar charge, this tail term is given by
\begin{equation}\label{eq:tail}
q^2 \int_{-\infty}^{\tau^-} \nabla_\mu G_{\rm ret} (z(\tau), z(\tau'))d\tau',
\end{equation}
where the world-line of the particle is given by $z(\tau)$, $q$ is the charge and $G_{\rm ret}(x,x')$ is the retarded Green function.
For electromagnetic and gravitational charges, the tail term involves a similar integral of the Green function along the past worldline.

The greatest difficulty in computing \eqref{eq:tail} arises from the computation of the Green function for points arbitrarily separated. For $x'\rightarrow x$, the Green function has a troublesome $\delta$-divergence behavior, while for $x$ and $x'$ separated by large distances, certain analytic expressions for the Green function are not valid. Within the matched expansion approach\cite{Poisson:Wiseman:1998},
these difficulties are avoided by splitting the integral \eqref{eq:tail} in two parts at a \emph{matching point} and tackling each part independently. We have given an example calculation of the \emph{distant past} contribution for the Nariai spacetime\cite{Casals:Dolan:Ottewill:Wardell:2009}. We focus here on the contribution from the recent past, the \emph{quasilocal} contribution.

Provided the quasilocal region is sufficiently small (within a \emph{normal neighborhood}), the retarded Green function is given by the Hadamard form,
\begin{equation}\label{eq:HadamardGF}
G_{\mathrm{ret}}(x,x') = \theta_{-} (x,x') \left\lbrace U (x,x') \delta \left( \sigma (x,x') \right) - V(x,x') \theta \left( - \sigma (x,x') \right) \right\rbrace,
\end{equation}
where $U(x,x')$ and $V(x,x')$ are regular, geometrical bitensors, $\delta$ is the Dirac-delta, $\theta_{-}$ is the step function and ensures causality and $\sigma(x,x')$ is the Synge world-function. Noting that $\sigma \neq 0$ for all points in the normal neighborhood contribution to \eqref{eq:tail}, it is clear that only $V(x,x')$ will contribute to the quasilocal part of \eqref{eq:tail}.

\section{Coordinate Expansion of the Green Function}
\begin{figure}
 \begin{center}
\begin{tabular}{m{6cm}m{5cm}m{0.75cm}}
 \psfig{file=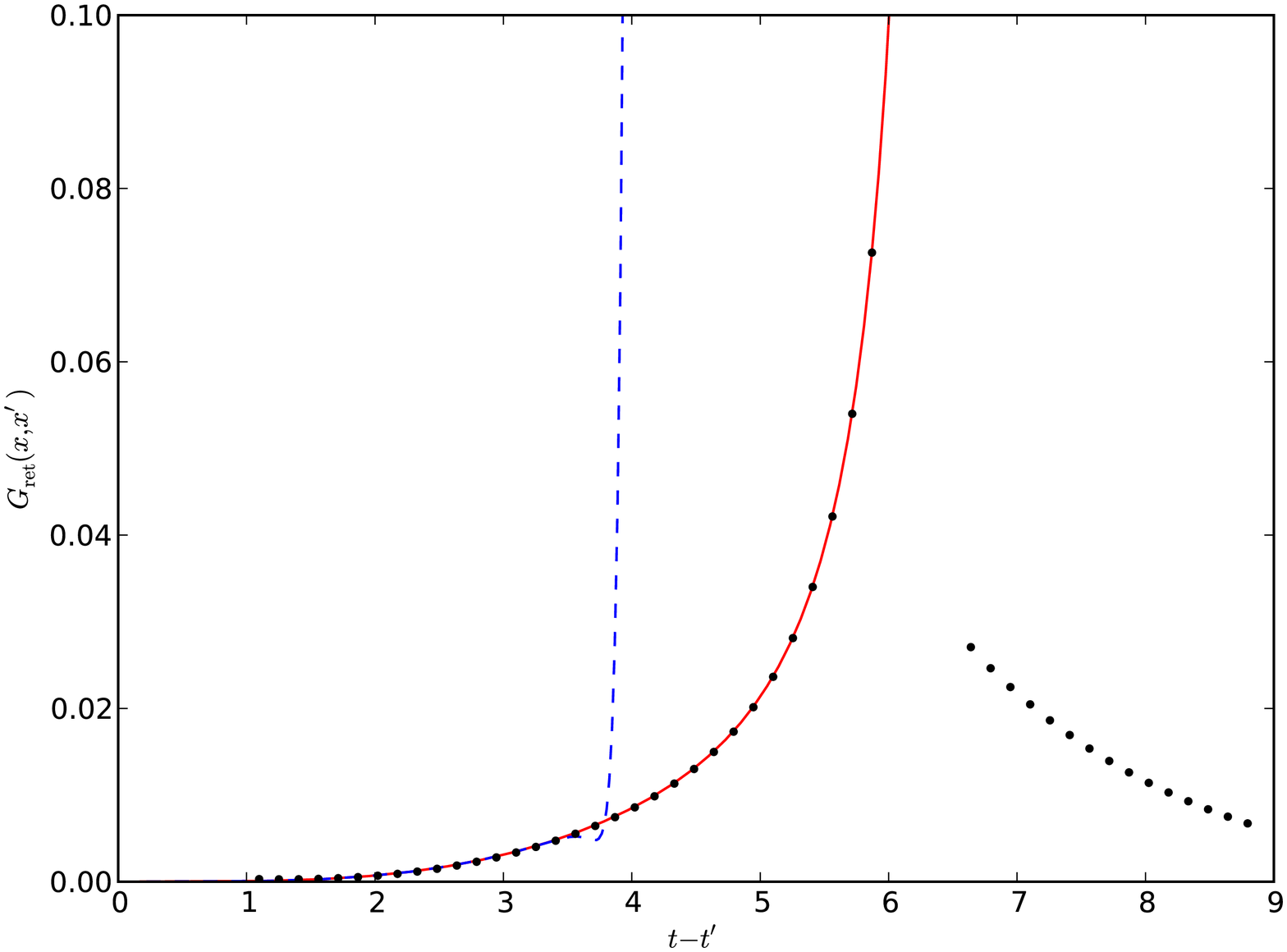,width=6cm,height=3.75cm} &
 \psfig{file=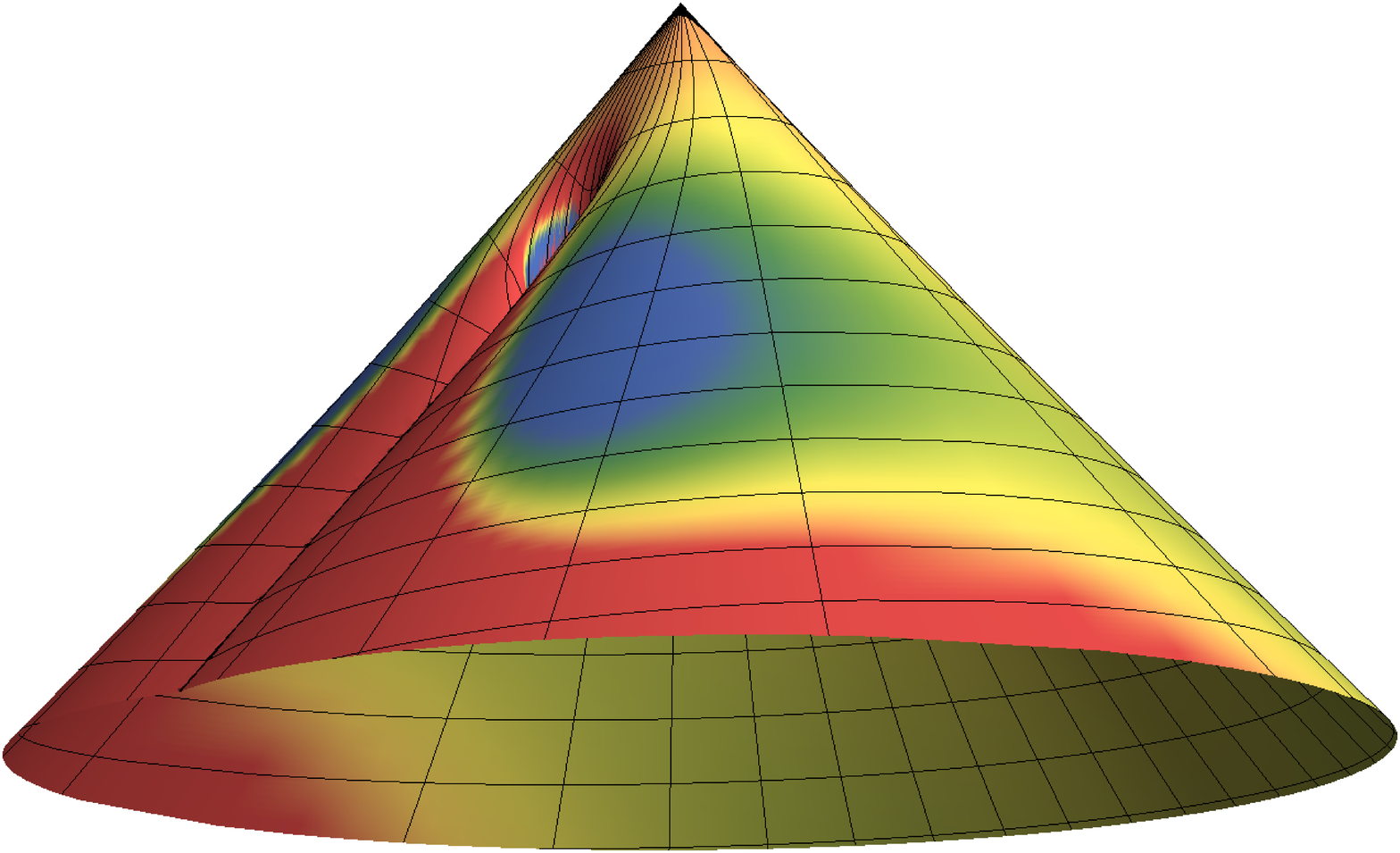,width=5cm} &
 \psfig{file=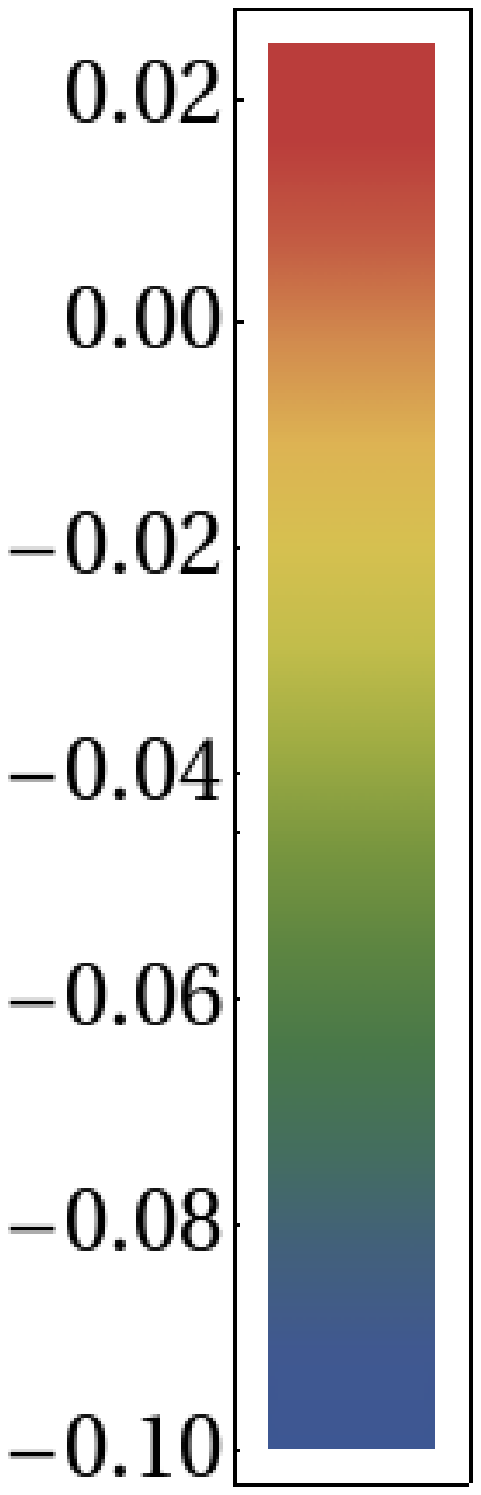,width=0.75cm}
\end{tabular}
\end{center}
\caption{
(Colour online) Left: The retarded Green function for points separated along the time direction in the Nariai spacetime. The regular Taylor series (blue dashed line) diverges before the edge of the normal neighborhood, which is at $t-t'=6.5$. The Pad\'e resummed series (solid red line) remains valid and closely matches the exact value (black dots) up to the edge of the normal neighborhood.
Right: \emph{$V(x,x')$ for a massless, scalar field along the light-cone of a point at $r=10M$ in Schwarzschild Spacetime.}
}
\label{fig:GF}
\end{figure}

The bitensor $V(x,x')$ 
may be written as an expansion in the coordinate separation of $x$ and $x'$. For static, spherically symmetric spacetimes, this takes the form
\begin{equation}
V(x,x') = \sum_{i,j,k=0}^{\infty} v_{ijk} (x) (t-t')^{2i}(\cos(\gamma)-1)^{j}(r-r')^{k}
\end{equation}
with $\gamma$ being the angular separation of the points. The coefficients $v_{ijk}(x)$ are calculated using methods which rely on the symmetry of the spacetime\cite{Anderson:2003}. We developed a \emph{Mathematica} code capable of calculating these coefficients to very high order (up to $(\Delta x^\alpha)^{50}$) and applied the theory of Pad\'e approximants to improve convergence and extend the domain of validity of the series\cite{QL}. This allowed us to accurately calculate (Fig.~\ref{fig:GF}) the Green throughout and up to the edge of the normal neighborhood, beyond which the Hadamard form can no longer be assumed to hold.

\section{Covariant Methods for Calculating the Green Function}
Covariant methods for calculating the Green function of the wave operator and the corresponding heat kernel, are central to a broad range of problems from radiation reaction to quantum field theory in curved spacetime
and quantum gravity. Traditionally, most attention has focused on the \textit{diagonal value}  of the heat kernel $K^{A}{}_{A}(x,x;s)$, corresponding to the coincidence limit $x'\rightarrow x$ of the Green function.  
By contrast, for the quasilocal contribution to the matched expansion self-force, we seek
expansions valid for $x$ and $x'$ as far apart as geometrical methods permit.   

An extremely elegant, non-recursive approach to the calculation of the Green function has been given by Avramidi~\cite{Avramidi:2000}, 
but as his motivation was to study the effective action in quantum gravity he was primarily interested in the coincidence limit. We presented\cite{Transport} 
Avramidi's approach in the language of transport equations and showed that it is ideal for numerical and symbolic computation. 
In so doing we built on the work of D\'ecanini and Folacci~\cite{Decanini:Folacci:2005a} who wrote many of the equations we presented and 
implemented them explicitly by hand. 
However, calculations by hand are 
quite impractical for the very high order expansions we would like for self-force calculations.
Instead, 
we used the transport equations as the basis for \textsl{Mathematica} code for algebraic calculations\cite{AvramidiCode} and \textsl{C} code for numerical calculations\cite{TransportCode}. 

Using our numerical code, we calculated $V(x,x')$ on the past light-cone of a point $x$ in the Schwarzschild spacetime. With $V(x,x')$ a solution of the homogeneous wave equation, $\Box V(x,x')=0$, this light-cone data could serve as initial data for a characteristic evolution of $V(x,x')$ \emph{within} the light-cone. For points within the normal neighborhood of $x$, there would be no caustics -- a frequent cause of difficulty in characteristic evolution codes. Nevertheless, the allure of evolving past caustics and beyond the normal neighborhood is appealing. It may reveal new insight into the relation between the Hadamard form and the structure of the Green function \emph{outside} the normal neighborhood.

\section{Acknowledgements}
BW has been supported by the Institute of Physics Research Student Conference Fund and by the Irish Research Council for Science, Engineering and Technology.


\bibliographystyle{ws-procs975x65}
\bibliography{main}

\begin{thebibliography}{10}

\bibitem{DeWitt:1960}
B.~S. DeWitt and R.~W. Brehme, {\em Ann. Phys.} {\bf 9}, 220 (1960).

\bibitem{Mino:Sasaki:Tanaka:1996}
Y.~Mino, M.~Sasaki and T.~Tanaka, {\em Phys. Rev.} {\bf D55}, 3457 (1997).

\bibitem{Quinn:Wald:1997}
T.~C. Quinn and R.~M. Wald, {\em Phys. Rev.} {\bf D56}, 3381 (1997).

\bibitem{Quinn:2000}
T.~C. Quinn, {\em Phys. Rev.} {\bf D62}, p. 064029 (2000).

\bibitem{Poisson:Wiseman:1998}
E.~Poisson and A.~G. Wiseman, Suggestion at the 1st {Capra Ranch} meeting on
  radiation reaction (1998).

\bibitem{Casals:Dolan:Ottewill:Wardell:2009}
M.~Casals, S.~Dolan, A.~C. Ottewill and B.~Wardell, {\em Phys. Rev.} {\bf D79},
  p. 124043 (2009).

\bibitem{Anderson:2003}
P.~R. Anderson and B.~L. Hu, {\em Phys. Rev.} {\bf D69}, p. 064039 (2004).

\bibitem{QL}
M.~Casals, S.~Dolan, A.~C. Ottewill and B.~Wardell, {\em Phys. Rev.} {\bf D79},
  p. 124044 (2009).

\bibitem{Avramidi:2000}
I.~G. Avramidi, {\em Heat Kernel and Quantum Gravity} (Springer, Berlin, 2000).

\bibitem{Transport}
A.~C. Ottewill and B.~Wardell (2009), arXiv:0906.0005.

\bibitem{Decanini:Folacci:2005a}
Y.~D\'ecanini and A.~Folacci, {\em Phys. Rev.} {\bf D73}, p. 044027 (2006).

\bibitem{AvramidiCode}
\url{http://www.barrywardell.net/research/code/avramidi}.

\bibitem{TransportCode}
\url{http://www.barrywardell.net/research/code/transport}.

\end{thebibliography}

\end{document}